\documentclass[conference]{IEEEtran}
\usepackage{amsmath,amssymb,amsfonts}
\usepackage{algorithmic}
\usepackage{graphicx}
\usepackage{textcomp}
\usepackage{xcolor}
\usepackage{booktabs}
\usepackage{float}
\usepackage{cite}
\usepackage[utf8]{inputenc}

\begin{document}
\title{Real-Time Control and Automation Framework for Acousto-Holographic Microscopy}

\author{
\IEEEauthorblockN{
Hasan Berkay Abdioğlu\textsuperscript{1},
Yağmur Işık\textsuperscript{1},
Mustafa İsmail İnal\textsuperscript{1},
Nehir Serin\textsuperscript{2}\\
Kerem Bayer\textsuperscript{1},
Muhammed Furkan Koşar\textsuperscript{1},
Taha Ünal\textsuperscript{3},
Hüseyin Uvet\textsuperscript{1}
}
\IEEEauthorblockA{\textsuperscript{1}Department of Mechatronics Engineering, Yildiz Technical University, Istanbul, Turkey}
\IEEEauthorblockA{\textsuperscript{2}Department of Control and Automation Engineering, Yildiz Technical University, Istanbul, Turkey\\
\IEEEauthorblockA{\textsuperscript{3}Department of Electronics and Communication Engineering, Yildiz Technical University, Istanbul, Turkey}
\textit{Corresponding author: huvet@yildiz.edu.tr}}
}

\maketitle

\begin{abstract}
Manual operation of microscopes for repetitive tasks in cell biology is a significant bottleneck, consuming invaluable expert time, and introducing human error. Automation is essential, and while Digital Holographic Microscopy (DHM) offers powerful, label-free quantitative phase imaging (QPI), its inherently noisy and low-contrast holograms make robust autofocus and object detection challenging. We present the design, integration, and validation of a fully automated closed-loop DHM system engineered for high-throughput mechanical characterization of biological cells. The system integrates automated serpentine scanning, real-time YOLO-based object detection, and a high-performance, multi-threaded software architecture using pinned memory and SPSC queues. This design enables the GPU-accelerated reconstruction pipeline to run fully in parallel with the 50 fps data acquisition, adding no sequential overhead. A key contribution is the validation of a robust, multi-stage holographic autofocus strategy; we demonstrate that a selected metric (based on a low-pass filter and standard deviation) provides reliable focusing for noisy holograms where conventional methods (e.g., Tenengrad, Laplacian) fail entirely. Performance analysis of the complete system identifies the 2.23-second autofocus operation-not reconstruction-as the primary throughput bottleneck, resulting in a 9.62-second analysis time per object. This work delivers a complete functional platform for autonomous DHM screening and provides a clear, data-driven path for future optimization, proposing a hybrid brightfield imaging modality to address current bottlenecks.
\end{abstract}


\section{Introduction}

Scientific progress, particularly in cell biology and materials science, is often built upon complex, repetitive experimental analysis. A significant portion of these tasks, such as sample screening, imaging, and focusing, is still performed manually by highly-skilled scientists and technicians. This manual operation is not only a source of significant human error and inter-operator bias, but it also consumes thousands of invaluable man-hours. This bottleneck ties high-knowledge personnel to routine labor, slowing the pace of discovery and representing an inefficient use of critical research funding. The automation of these experiments is therefore an essential step, as it promises to eliminate human bias, ensure experimental consistency, and, most importantly, free expert researchers to focus on data analysis and hypothesis-driven science \cite{Freedman2015}. 

Digital Holographic Microscopy (DHM) is a prime candidate for such automation. As a powerful quantitative phase imaging (QPI) technique \cite{Masters2012}, it provides rich, label-free, quantitative data, enabling the measurement of cell morphology and mechanical properties from a single acquisition. However, the operation of DHM systems and the analysis of the resulting off-axis holograms are often complex, requiring manual focusing and intervention.

While automated microscopy systems are common, many rely on clean, high-contrast brightfield or fluorescence imaging. The off-axis holograms used in QPI, however, are inherently noisy and present low-contrast features. This makes the robust, unmanned application of standard computer vision algorithms for object detection and autofocus extremely challenging. Most existing systems fail to provide a complete, closed-loop solution that integrates autonomous scanning, intelligent holographic focusing, and a high-performance acquisition and reconstruction pipeline for mechanical analysis.

In this work, we present the design and validation of a fully automated system built around a bespoke, custom-built DHM (off-axis interferometer) \cite{Varol2022}, engineered for high-throughput mechanical characterization. Our system integrates (A) a robust multi-threaded software architecture built on pinned memory, SPSC queues, and GPU-accelerated processing; (B) a serpentine scanning stage with a novel adaptive logic for boundary handling and duplicate rejection; (C) a real-time YOLO-based object detector; and (D) a multi-stage holographic autofocus strategy that overcomes the limitations of noisy interferograms.

This paper details the complete methodology for this integrated system, from its opto-mechanical components to its parallel software architecture. We then provide a rigorous validation of its key components, presenting a data-driven analysis of the autofocus performance, detection speed, and overall system throughput. We conclude with a critical discussion of the system's performance, identifying the primary bottlenecks and proposing a clear, data-driven path for future optimization.

\section{Methodology}

The methodology details the design and integration of an automated Digital Holographic Microscopy (DHM) system engineered for real-time, high-throughput detection, and topographical mapping and mechanical characterization (Young's Modulus) of biological cells. The system is partitioned into four primary subsystems: \ref{sec:optic} the interferometric imaging subsystem for hologram acquisition; \ref{sec:motor} the automated scanning and positioning subsystem for sample manipulation; \ref{sec:control} the real-time image processing and control subsystem for object detection and automated focus; and \ref{sec:software} the real-time reconstruction subsystem for quantitative phase retrieval.

The system architecture integrates optical, mechanical, and computational components into a closed-loop platform, managed by a central host computer. The operational workflow is designed for automated, multi-stage analysis at each scanning position.

First, the host computer directs the automated scanning subsystem to position the sample, which consists of a PDMS chip with an integrated piezoelectric (PZT) transducer. The imaging subsystem acquires a single survey hologram, which is immediately processed by the control subsystem to perform object detection and per-object autofocus.

Once a target is identified and focused, the system proceeds to the mechanical analysis phase. While the integrated PZT provides continuous 10 Hz excitation, the imaging subsystem is triggered to acquire a rapid burst of 250 holograms at 50 frames per second (fps), capturing the sample's dynamic response.

This high-frequency holographic sequence is then made available to the real-time reconstruction subsystem, which processes the acquired holograms to compute the quantitative phase (height) map. This phase data is subsequently used as an input to the Hertzian elasticity model \cite{Mishra2014} to derive the object's mechanical stiffness (Young's Modulus).

\subsection{Interferometric Imaging Subsystem} \label{sec:optic}

The optical core is a custom-built, off-axis DHM in a Mach-Zehnder interferometer configuration. A coherent, solid-state green laser (532 nm) serves as the illumination source. The beam is split into an object arm, which illuminates the sample, and a reference arm. The two beams are recombined at a slight angle to generate the off-axis hologram, which is recorded directly by a Basler BOA monochrome CMOS sensor in a lensless configuration.

\begin{figure}[h!]
    \centering
    \includegraphics[width=1\linewidth]{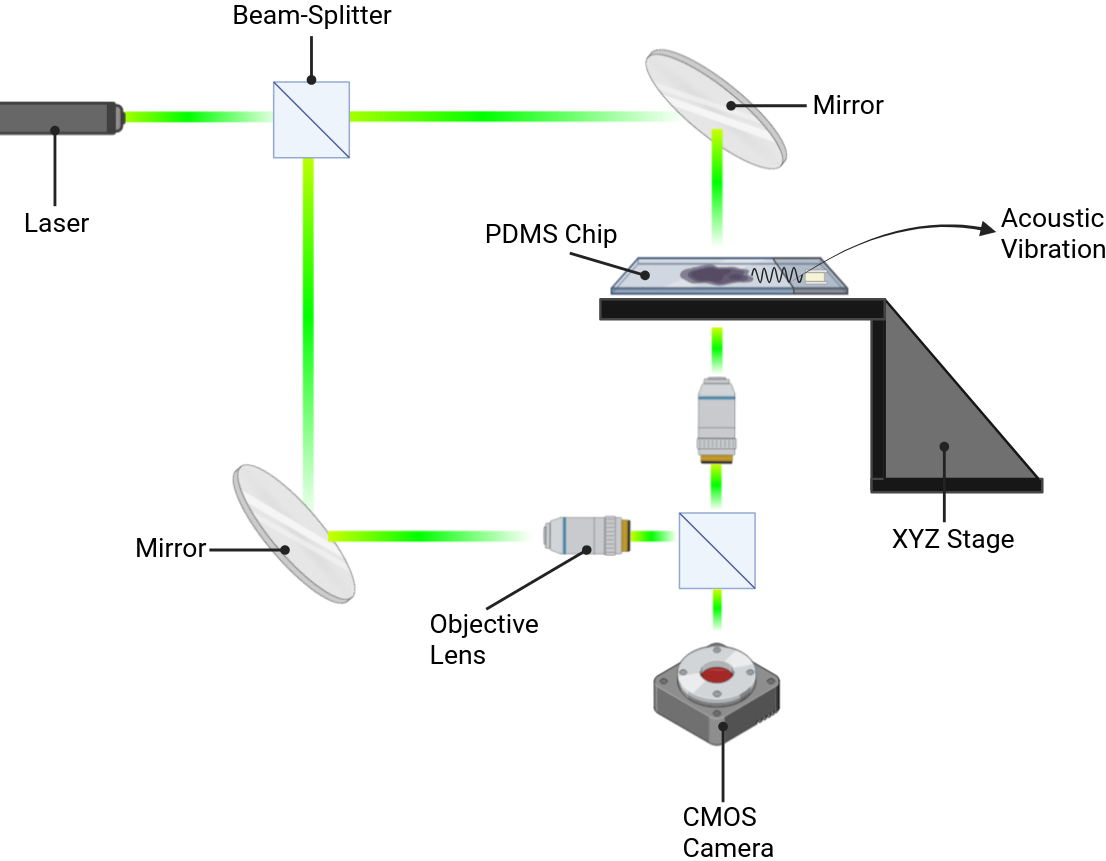}
    \caption{Side-view of the Mach-Zehnder Interferometer Setup}
\end{figure}
\subsection{System Architecture}

\subsection{Automated Scanning and Positioning Subsystem} \label{sec:motor}

Automated sample traversal is achieved using a high-precision, three-axis (XYZ) stepper motor stage assembly, managed by a PI (Physik Instrumente) Mikromove C-884 controller.

The system is configured to scan a user-defined rectangular area, delimited by $x_{min}$, $x_{max}$, $y_{min}$, and $y_{max}$ coordinates, allowing for dynamic region selection. To maximize throughput, the stage executes a serpentine scanning pattern. The X-axis first traverses from $x_{min}$ to $x_{max}$. Upon completion of this sweep, the Y-axis increments by a single step. The X-axis then traverses in the reverse direction ($x_{max}$ to $x_{min}$). This path is repeated until the Y-axis reaches $y_{max}$, minimizing the mechanical overhead associated with stage acceleration and deceleration.

The incremental step size in X and Y is calculated based on the optical Field of View (FOV), which is determined by the CMOS sensor's active area and the system's optical magnification. The step size is configured to be slightly less than the FOV to ensure sufficient overlap between adjacent fields to minimize object misses between the boundaries.

\subsection{Real-Time Image Processing and Control Subsystem} \label{sec:control}

This subsystem is responsible for the real-time, closed-loop control of the acquisition process. It executes a multi-stage detection and focusing pipeline at each scanning position and implements adaptive logic to handle boundary-case objects, ensuring both comprehensive and non-redundant data capture.

A core component of this subsystem is a custom autofocus algorithm, as conventional methods (e.g., gradient, statistical, energy-based) were found to be unreliable for our off-axis holograms. Our focusing metric is defined as a function $f(Z)$, where $Z$ is the position of the Z-axis stage. This metric is computed by applying a low-pass filter to the hologram and then calculating the standard deviation of the result \cite{Malik2020}. The resulting metric is unimodal with respect to $Z$, and its minimum, $argmin(f(Z))$, corresponds to the optimal focal plane. To find this minimum efficiently with the fewest stage movements, we employ Brent's method for numerical minimization.

The primary processing pipeline at each discrete (x,y) stage position is as follows:


\begin{figure}[!t]
    \centering
    \includegraphics[height=0.7\textheight, keepaspectratio]{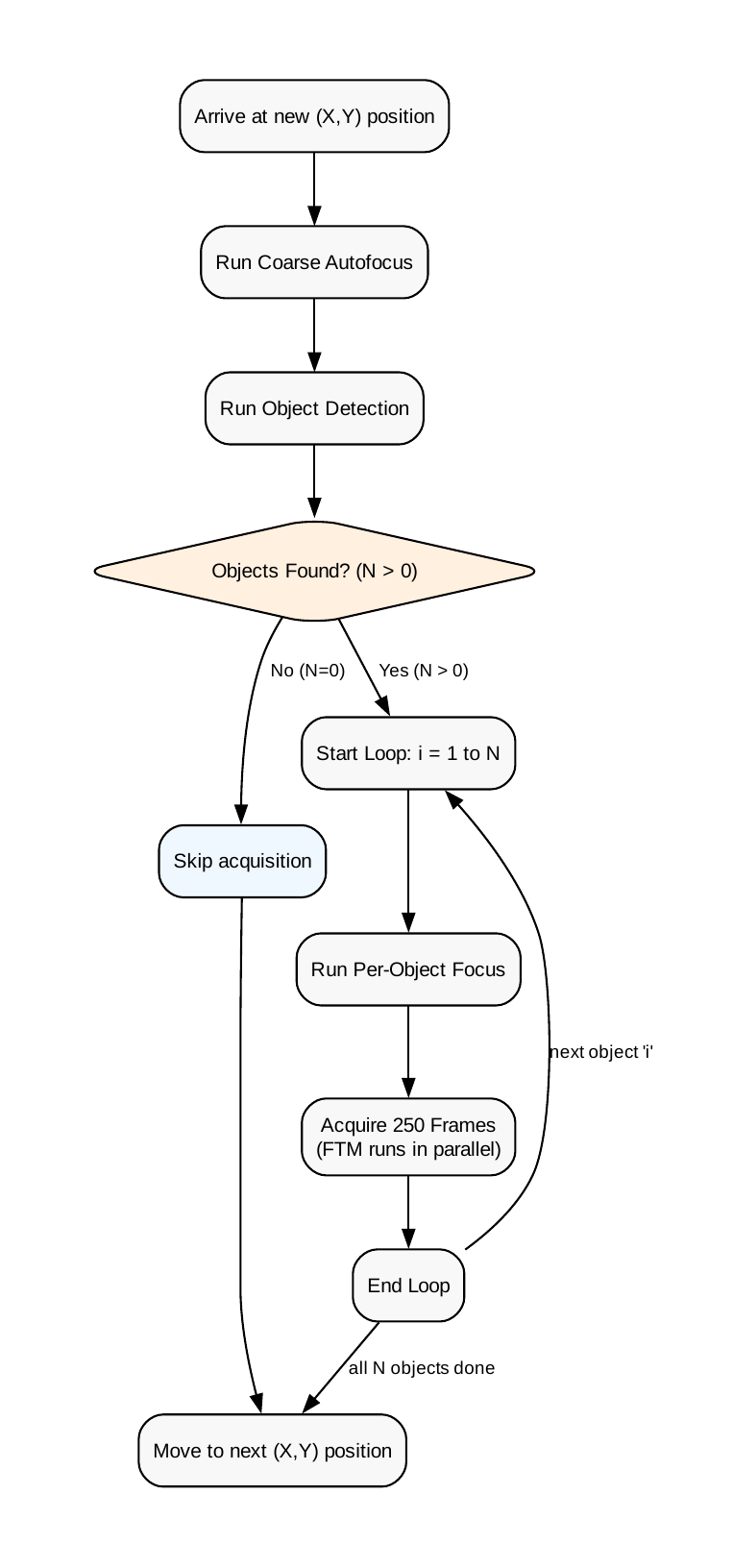}
    \caption{Process Chart for the Real-Time Image Processing and Control Subsystem.}
    \label{fig:control_flow}
\end{figure}

\begin{enumerate}
    \item Initial Coarse Autofocus: The autofocus algorithm is first applied to the entire FOV. This step establishes a baseline focus sufficient to optimize the performance of the subsequent detection model.

    \item Object Detection: The frame is processed by a high-speed YOLOv11 Nano model \cite{Jocher_Ultralytics_YOLO_2023}. This lightweight deep learning model identifies and localizes all objects of interest (cells), returning their bounding boxes.
    \item Per-Object Fine Autofocus: To compensate for Z-axis variations within the current field of view, the system iterates through each detected object. For each object, the autofocus metric is re-computed exclusively within its bounding box (ROI), and the Z-axis is adjusted to this new, object-specific focal plane.

    \item Sequential Acquisition: Once fine focus is achieved for a target object, the system triggers the acquisition of the 250-hologram sequence at 50 fps.
\end{enumerate}

A critical function of this subsystem is the management of objects identified at the FOV boundaries. If a detected bounding box intersects the camera FOV boundaries, the primary scanning loop is paused. The stage controller is commanded to move in sub-increments to re-center that specific object, ensuring its complete capture. Furthermore, to prevent redundant acquisition in overlapping scan areas, a spatial hash map is maintained. Upon acquiring an object, its absolute world coordinates (calculated by combining the high-precision stage position with the object's pixel-based coordinates) are stored. Newly detected objects are checked against this map, and those within a defined spatial tolerance of a recorded entry are skipped

\subsection{Real-Time Reconstruction and Quantification Subsystem} \label{sec:software}

This subsystem performs the final computational steps, converting the 250-frame holographic sequences into quantitative mechanical properties.

The reconstruction pipeline is based on the Fourier transform method (spatial filtering) \cite{Takeda1982}. For each hologram, a 2D Fast Fourier Transform (FFT) is computed, which spatially separates the zero-order and twin-image terms in the frequency domain. The center of the positive side-band is located, and a circular filter is applied to isolate this spectrum. This isolated spectrum is then computationally shifted (rolled) to the center of the k-space, and an inverse FFT is applied to retrieve the complex wavefront, $U(x,y)$. The phase map is extracted by computing the argument of the complex field. A phase unwrapping algorithm is then applied, and the result is multiplied by a system-specific calibration constant to yield the final quantitative phase map, representing the Optical Path Difference (OPD), or height map.

This entire reconstruction process is repeated for all 250 frames in the sequence, yielding a time-resolved height map sequence. This dynamic data, which captures the cell's vibration induced by the 10 Hz excitation, is then processed as a complete set. Once all 250 frames are available, the data is used as input to the Hertzian elasticity model \cite{Mishra2014} to compute the mechanical stiffness (Young's Modulus) of the object.

This high-throughput computation is enabled by a multi-threaded software architecture orchestrated by the main thread. The design is partitioned into three dedicated threads: a producer (camera acquisition), a consumer (reconstruction), and a display thread.

\begin{figure}[h!]
    \centering
    \includegraphics[width=1\linewidth]{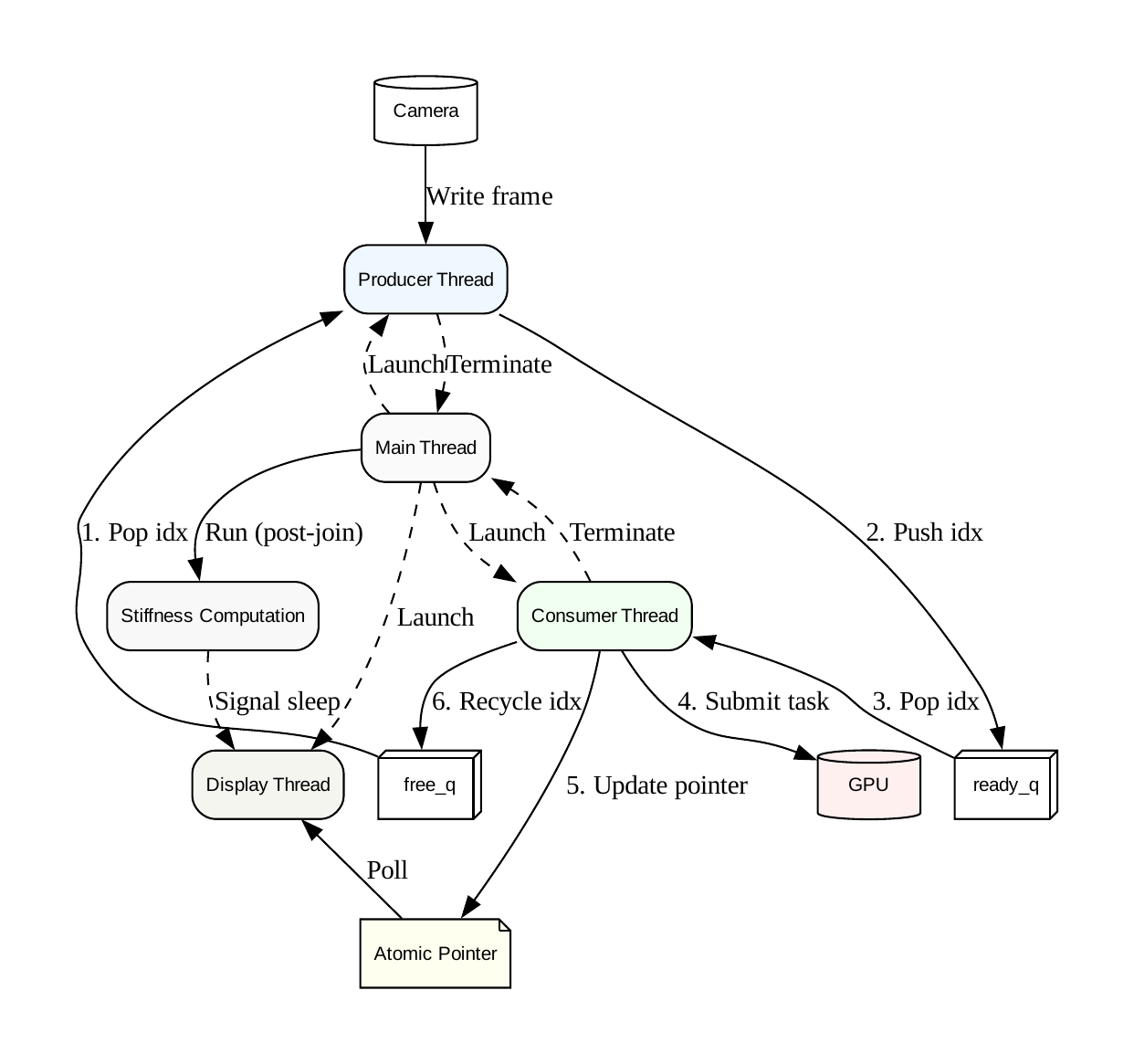}
    \caption{Data Flow Diagram for the Real-Time Reconstruction and Quantification Subsystem}
\end{figure}

\begin{figure*}[!t]
    \centering
    \includegraphics[width=\textwidth]{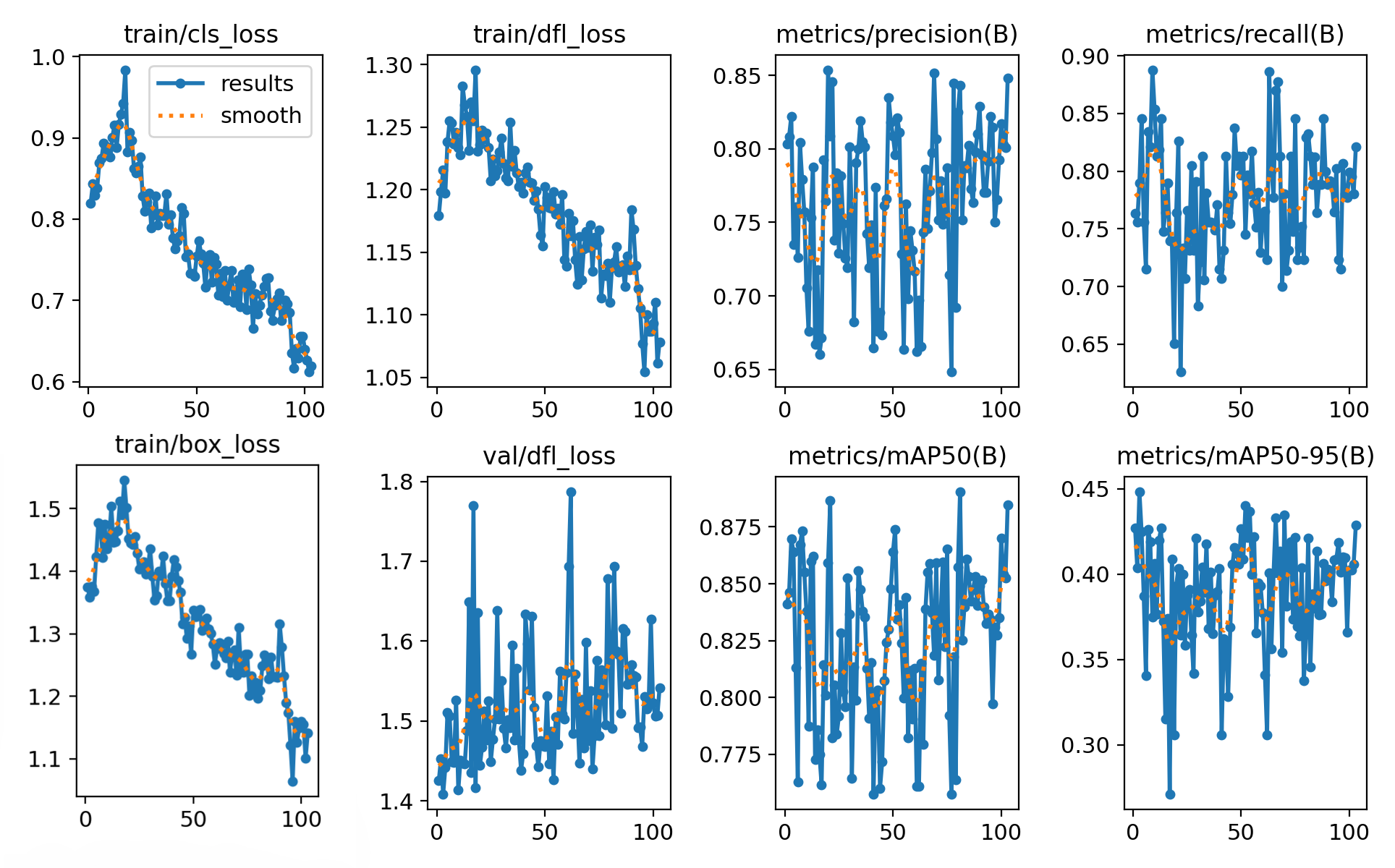}
    \caption{Training and evaluation metrics of the YOLO model across 100 epochs.}
    \label{fig:od_result}
\end{figure*}
\begin{itemize}
    \item Memory and Queue Architecture: At initialization, a large virtual memory arena is pre-allocated as page-locked (pinned) host memory. This strategy reserves a contiguous address space for all frame buffers, a requisite for enabling high-bandwidth, asynchronous Direct Memory Access (DMA) transfers \cite{NVIDIA2015} between the host system and the compute device (GPU) . Data flow is managed by two lock-free SPSC (Single Producer Single Consumer) queues, a free queue and a ready queue, which handle the circulation of frame buffer indices.

    \item Producer Thread: The producer thread is responsible for data acquisition. It pops an index from the free queue, writes the acquired hologram data into the corresponding buffer, and pushes that index onto the ready queue. This is repeated until all 250 frames are acquired, at which point the thread terminates.

    \item Consumer Thread: The consumer thread orchestrates the asynchronous GPU reconstruction. It dequeues frame indices from the ready queue and submits the full FTM computation (FFT, filtering, unwrapping) to a double-buffered pipeline using hardware streams. This non-blocking thread polls for completion events; upon detecting a finished frame, it updates an atomic pointer (for the display) and enqueues the processed index onto the free queue for recycling. The thread terminates once the producer has finished and all queued computations are complete.

    \item Display Thread: The display thread operates asynchronously. It continuously polls (spies on) the atomic pointer and renders the data it references. This shows the real-time height maps as they are computed. This thread also monitors a signal to enter a wait (sleep) state.

    \item Main Thread: The main thread orchestrates this entire process. It first launches the producer, consumer, and display threads. It then waits for both the producer and consumer threads to complete. Once joined, the main thread is assured that all 250 height maps are computed. It then proceeds to compute the final Young's Modulus from the full 250-frame sequence. After this calculation, it signals the display thread to enter its wait state, thus completing the analysis cycle for the current object.
\end{itemize}

\section{Results}

This section presents the experimental results that validate the design and performance of the key components of the automated microscopy system. The validation is presented in four parts: \ref{sec:od} we quantify the accuracy and inference speed of the object detection subsystem; \ref{sec:autofocus} we provide a comparative analysis of our novel autofocus algorithm against conventional methods to demonstrate its reliability and efficiency; \ref{sec:rt_perf} we characterize the real-time software architecture's performance, including processing throughput and acquisition-to-display latency; and \ref{sec:overall_perf} we summarize the overall system throughput by presenting a complete breakdown of the automated analysis pipeline.

\subsection{Object Detection Performance} \label{sec:od}

 The detection accuracy of the YOLO-based object detection model trained to identify target structures (e.g., cells, beads) in microscopic imaging set-up was evaluated as first thing. The evaluation includes both quantitative metrics and qualitative visualizations to assess detection accuracy, localization precision, and robustness under varying imaging conditions. Following this, additional components of the system—such as dynamic exposure control, focus optimization, and scanning accuracy—are evaluated to demonstrate the integrated system’s effectiveness in a practical imaging context.

The performance of the YOLOv11 model was evaluated using standard object detection metrics. As shown in Figure \ref{fig:od_result}, all loss components converge stably, indicating effective learning and reduced overfitting. The precision-recall (PR) curve reflects the model’s ability to maintain high precision even at higher recall thresholds.

The figure also shows that the model achieved a mean Average Precision (mAP) of 0.85 at an IoU threshold of 0.5 (mAP@0.5), with precision and recall values of 0.80 and 0.83, respectively—resulting in a strong detection F1-score.

\begin{figure}[h!]
   \centering
   \includegraphics[width=0.48\textwidth]{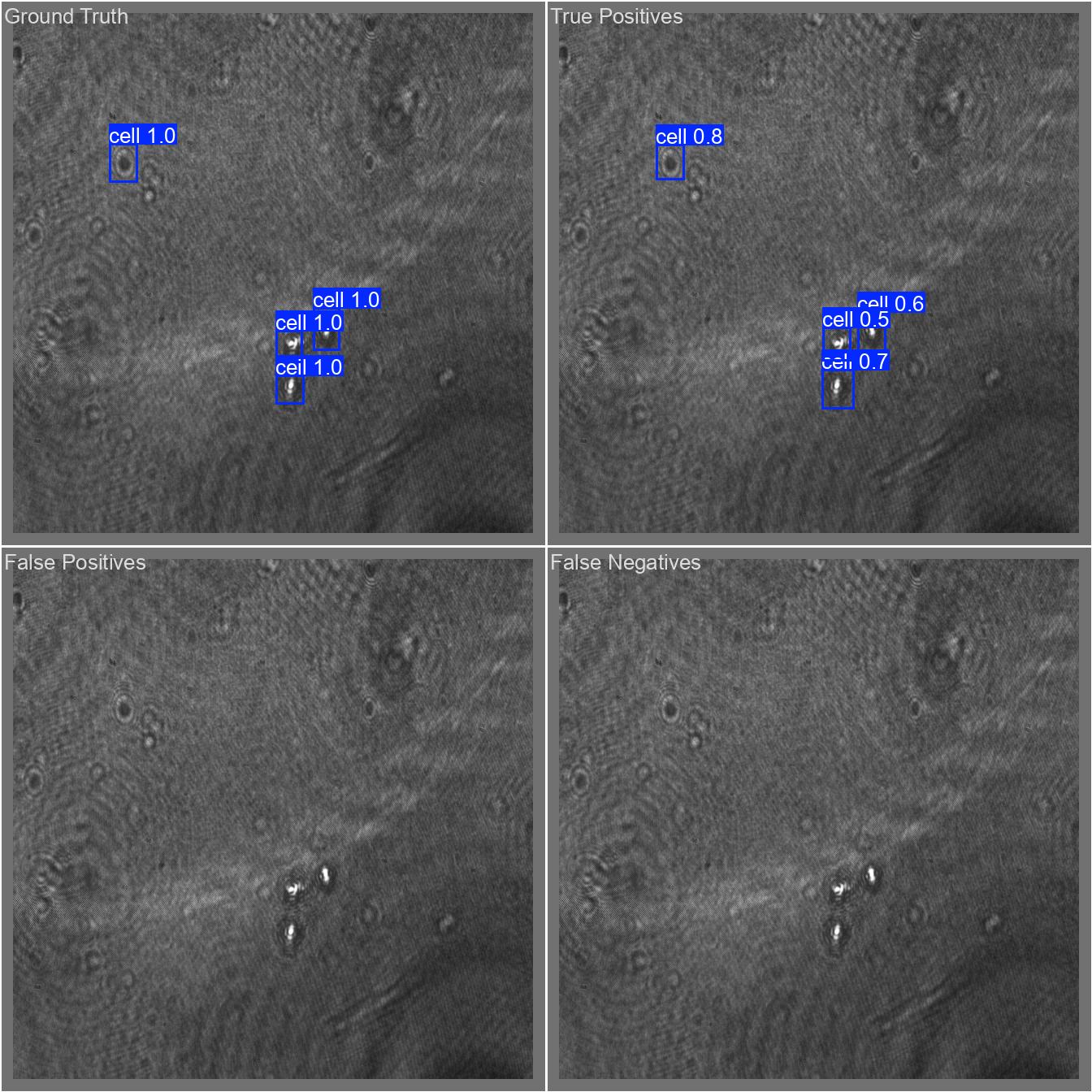} 
   \caption{
   Sample inference result showing ground truth and predicted bounding boxes. True positives, false positives and false negatives are separately remarked in the results.}
   \label{fig:detection}
\end{figure}

To further evaluate the detection performance, a representative inference result is shown in Figure \ref{fig:detection}, illustrating the predicted bounding boxes compared to the ground truth annotations.

An example detection alongside the ground truth, illustrating true positive, false positive, and false negative results. The model effectively identifies the majority of target structures while accurately excluding shadow projections that resemble cellular structures is presented.

This result demonstrates the model’s robust detection capability in distinguishing true target structures from visually similar artifacts, such as shadow projections. By effectively minimizing false positives and false negatives, the model shows promise for reliable application in automated cell detection tasks, which is critical for accurate biological analysis and diagnostic processes.

\subsection{Autofocus Algorithm} \label{sec:autofocus}

This subsection validates the performance of the selected autofocus strategy. The validation is performed in two parts: first, we demonstrate the reliability of the chosen focus metric by benchmarking it against conventional methods. Second, we validate the efficiency of the numerical search algorithm used to find the minimum.

To create a ground-truth dataset for this benchmark, a Z-stack of 500 images was acquired with a fine increment of 0.5 um per step. The optimal focal index within this stack was determined by manual inspection.

The selected focus metric was benchmarked against five conventional methods: Tenengrad, Laplacian, variance, normalized variance, and entropy.

\begin{figure}[h!]
    \centering
    \includegraphics[width=\linewidth]{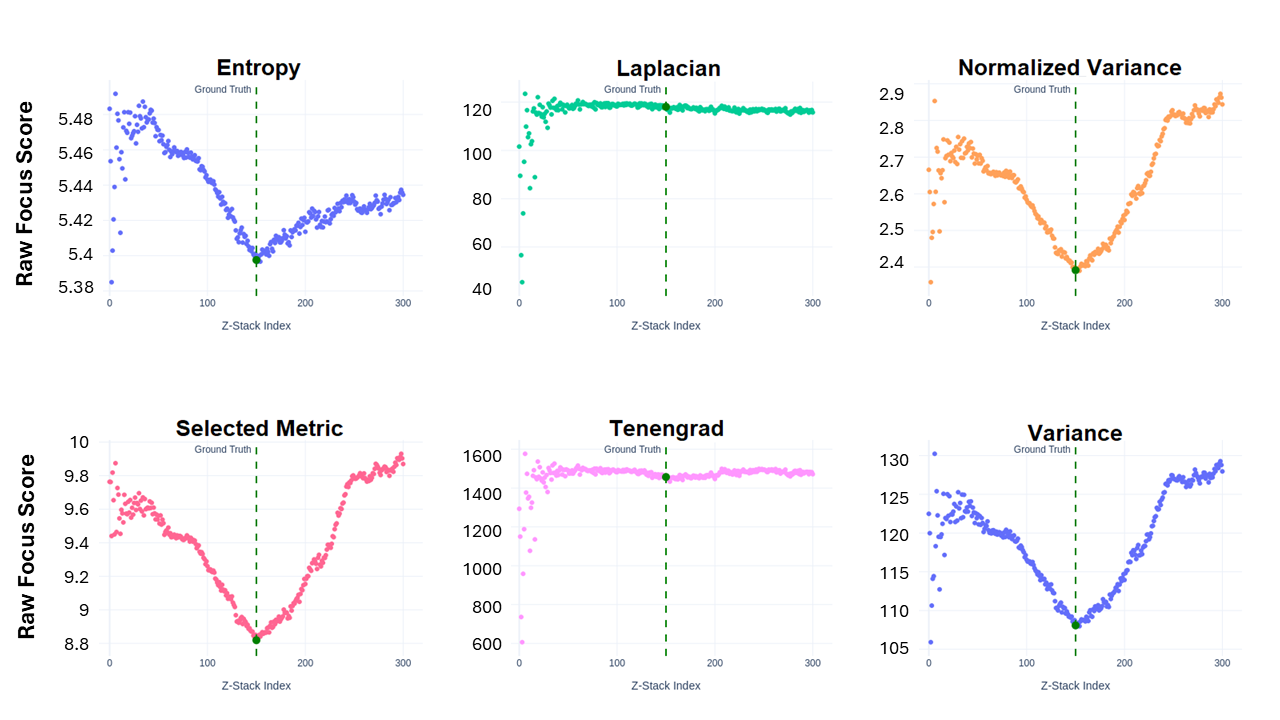}
    \caption{Best Case Scenario of the Methods}
    \label{fig:best_case}
\end{figure}

\begin{figure}[h!]
    \centering
    \includegraphics[width=\linewidth]{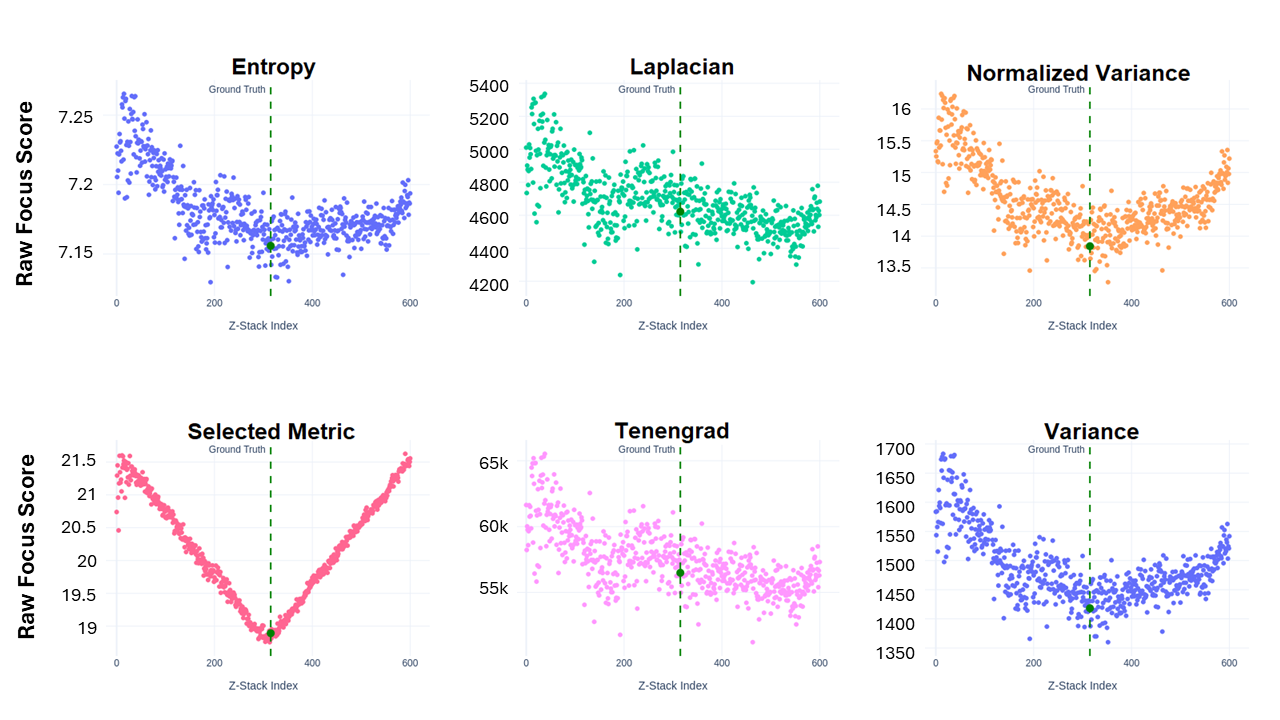}
    \caption{Worst Case Scenario of the Methods}
    \label{fig:worst_case}
\end{figure}

The qualitative performance of these metrics is illustrated in Figure \ref{fig:best_case} and Figure \ref{fig:worst_case}.

Figure \ref{fig:best_case} shows a best-case scenario with a high-contrast, clearly defined object. Even in this ideal case, gradient-based methods (Tenengrad, Laplacian) fail, exhibiting erratic behavior. While other methods (variance, entropy) find the general focal region, the selected metric provides a  smoother curve with substantially less noise near the minimum, making it a more reliable target for numerical minimization.

Figure \ref{fig:worst_case} shows a challenging worst-case scenario (e.g., low-contrast object, high background noise). Here, all conventional methods fail entirely. Their response curves are dominated by noise, making it impossible to distinguish a true minimum. The selected metric, however, remains robust, providing a clear and unimodal curve, demonstrating its superior reliability in non-ideal conditions.

A quantitative comparison of the metrics' reliability and computational cost, averaged over 50 different Z-stacks, is presented in Table \ref{tab:focus_metrics}.

\begin{table}[h!]
\centering
\caption{Focus Metric Performance. Error is reported relative to the ground-truth focal index.}
\label{tab:focus_metrics}
\begin{tabular}{lrr}
\toprule
Metric & Error (Mean $\pm$ SD) & Time (Mean $\pm$ SD) [ms]\\
\midrule
Selected Method &           -4.29 $\pm$ 12.74 &                6249.49 $\pm$ 5953.66 \\
                 Variance &          103.71 $\pm$ 85.84 &                  173.02 $\pm$ 120.41 \\
           Norm. Variance &          104.14 $\pm$ 85.22 &                  222.16 $\pm$ 155.50 \\
                Tenengrad &         108.29 $\pm$ 126.79 &                  865.65 $\pm$ 735.26 \\
                Laplacian &         140.86 $\pm$ 102.15 &                  482.08 $\pm$ 337.94 \\
                  Entropy &         145.29 $\pm$ 107.70 &                    81.47 $\pm$ 39.09 \\
\bottomrule
\end{tabular}
\end{table}

As shown quantitatively in Table \ref{tab:focus_metrics}, the selected std-dev (low-pass) metric exhibited the lowest error and standard deviation, confirming its superior accuracy and consistency. While it has a substantial execution time, this trade-off is necessary as conventional methods fail entirely in non-optimal scenarios (Figure \ref{fig:worst_case}), making them unreliable for robust, automated operation.

We then validated the real-world performance of Brent's method in finding the minimum of the robust std-dev (low-pass) metric. As summarized in Table \ref{tab:brent_perf}, the algorithm's performance is characterized by high efficiency and low bias. The method required an average of only 8 queries to find a minimum.

Critically, the algorithm demonstrates a 42.8\% success rate, defined as locating the exact (zero-tolerance) focal index within the 500-step Z-stack. While the standard deviation of the final error is high, this is strongly offset by an extremely low mean error, indicating a lack of systematic bias. The combination of a high, zero-tolerance success rate, low mean error, and minimal queries confirms its effectiveness for our automated system.

\begin{table}[h!]
\centering
\caption{Minimization Algorithm Performance. Error is given as mean $\pm$ standard deviation. 'Mean Queries' is the average number of steps to find a minimum.}
\label{tab:brent_perf}
\begin{tabular}{lrrr}
\toprule
Algorithm & Error & Success Rate & Mean Queries \\
\midrule
Brent's Method & -0.28 $\pm$ 23.07 & 42.8 \% & 8.0 \\
\bottomrule
\end{tabular}
\end{table}

\subsection{Real-Time Acquisition and Visualization Performance} \label{sec:rt_perf}

The performance of the multi-threaded software architecture was validated to ensure its capability for high-throughput, real-time data handling. The performance of each thread was confirmed:

\begin{itemize}
    \item Producer Thread: The producer's throughput is governed by the native camera speed of 50 fps. The page-locked memory arena and lock-free SPSC queue are designed to service this 50 Hz data stream at video-rate without data loss, ensuring the acquisition is not a bottleneck.

    \item Consumer Thread: The consumer thread offloads all FTM reconstruction to the GPU. As established in the literature, GPU-accelerated FTM pipelines can process megapixel-sized holograms at rates (e.g., 35-129 fps \cite{Backoach2016}) that comfortably exceed our 50 fps acquisition speed. This confirms the reconstruction process is not a bottleneck and can operate in real-time.

    \item Display Thread: The asynchronous display thread was benchmarked in our implementation. While it achieves a fluid ~30 fps for 512x512 images, the rate for megapixel-level images is lower, at ~13 fps for 1024x1024 images. Critically, this does not affect system performance. Due to its asynchronous, non-blocking design, the display thread acts purely as an observer. It does not introduce any latency, back-pressure, or stalls into the primary acquisition and reconstruction pipeline.
\end{itemize}

\subsection{Overall System Performance} \label{sec:overall_perf}

The system's total sequential processing time, $T_{total}$, for a single field of view (FOV) is a variable dependent on $N$, the number of objects detected within that FOV. The pipeline's logic and component timings are presented in Table \ref{tab:component_times}.

\begin{table}[h!]
\centering
\caption{Measured average execution times for the sequential components of the analysis pipeline.}
\label{tab:component_times}
\begin{tabular}{lr}
\toprule
\textbf{Component} & \textbf{Time (seconds)} \\
\midrule
Coarse Autofocus ($T_{coarse}$) & 2.23 $\pm$ 0.41 \\
Object Detection ($T_{detect}$) & 0.16 $\pm$ 0.01 \\
Per-Object Focus ($T_{fine}$) & 2.23 $\pm$ 0.41 \\
Hologram Acquisition ($T_{acq}$) & 5.0 (Constant) \\
\bottomrule
\end{tabular}
\end{table}

The total sequential time for one FOV follows two distinct paths:

\begin{enumerate}
    \item Empty FOV (N=0): If no objects are detected, the system performs the coarse focus and detection, then immediately moves to the next position.
        $$ T_{total} = T_{coarse} + T_{detect} $$

    \item Non-Empty FOV (N\textgreater 0): If one or more objects are detected, the system performs the initial steps, then iterates N times, executing a fine-focus and a 5-second acquisition for each object. As established in the previous section, the FTM reconstruction is performed in parallel by the consumer thread during this 5-second acquisition, adding no sequential overhead.
        $$ T_{total} = T_{coarse} + T_{detect} + N \times \left( T_{fine} + T_{acq} \right) $$
\end{enumerate}

This model accurately reflects the system's throughput. An empty FOV is efficiently skipped in 2.39 seconds, while a FOV with a single object is fully processed in 9.62 seconds. This object-density-dependent throughput validates the system's autonomous operation and provides a clear model of its performance.

\section{Discussion}

This work presents the successful design, integration, and validation of a fully automated Digital Holographic Microscopy (DHM) system, which autonomously performs sample scanning, object detection, multi-stage focusing, and high-speed, GPU-accelerated acquisition for mechanical analysis. Our results confirm that all hardware and software subsystems operate cohesively in a closed loop, and the resulting performance model validates its utility for high-throughput experiments. The system's primary strengths lie in its robust software-driven automation, the confirmed reliability of the chosen autofocus metric, and a high-performance, parallelized software architecture that ensures reconstruction does not bottleneck the acquisition.

Despite these successes, the project's focus on software-centric solutions and high-speed acquisition, while neglecting potential hardware or setup modifications, led to significant compromises in algorithm selection and overall system efficiency. The system's primary limitation, as quantified in the Overall System Throughput model, is the 2.23-second autofocus operation, which is the dominant bottleneck. This is not a limitation of the motor, but a direct consequence of the computationally expensive focus metric required to find a reliable minimum in noisy, low-contrast off-axis holograms.

This reliance on a complex software solution stems from a larger methodological compromise. By adhering strictly to a holographic-only imaging modality, we were forced to accept sub-optimal algorithms. The high-speed YOLOv11 Nano model, for example, was chosen due to an over-emphasis on inference speed, but it produces unneccesary error rates and requires extensive, setup-specific training. It struggles significantly with noise and setup variations, making the system brittle. Similarly, the computationally expensive std-dev (low-pass) metric was the only method found to be reliable for autofocus, solving the focusing problem but at the cost of creating the system's single largest performance bottleneck.

The system's performance and robustness could be dramatically improved by addressing these limitations, primarily by incorporating a hybrid imaging modality. By integrating a software-controlled shutter into the interferometer's reference arm, the system could optionally block the reference beam to capture standard brightfield micrographs. This single hardware modification would unlock a cascade of superior algorithmic solutions.

A brightfield-enabled system would almost certainly obviate the need for the slow focus metric; a simple, fast metric like variance would likely become viable, reducing the focus time to be limited only by the motor's movement. Furthermore, with clean brightfield images, we could replace the error-prone YOLO model with a state-of-the-art, pre-trained segmentation model like Cellpose \cite{Stringer2021}. This would provide vastly superior accuracy and segmentation masks while making the system more general and robust. The minor additional runtime overhead of such models (milliseconds) would be entirely negligible when compared to the 2.23-second autofocus and 5-second acquisition bottlenecks, making the initial choice of YOLO an unnecessary over-optimization.

An alternative software-centric approach would be to implement an Extended Depth of Focus (EDOF) algorithm \cite{Ferraro:05}. A sparse Z-stack could be acquired at each FOV, fused into an all-in-focus image, and then processed to identify all objects and their focal planes in a single pass. This carries its own trade-offs, such as the high-time penalty for processing empty FOVs, but it represents another valid path for future optimization. Finally, while the asynchronous display thread's performance was not a focal point, its inability to maintain video-rate at high resolutions could be improved, though it does not impact the system's core acquisition performance.

\section{Conclusion}

We have presented the design, integration, and performance validation of a fully automated Digital Holographic Microscopy (DHM) system, engineered for autonomous mechanical characterization. The system successfully integrates automated serpentine scanning, real-time object detection, a robust multi-stage autofocus, and a high-performance, GPU-accelerated parallel processing pipeline.

Our key contribution is the demonstration of this complete, closed-loop framework, which frees high-knowledge personnel from the time-intensive, manual operation of the microscope. This automation not only eliminates human error and ensures experimental consistency, but also enables high-throughput data acquisition. We validated that the computationally-intensive FTM reconstruction runs entirely in parallel with the data acquisition, adding no sequential overhead.

However, our analysis identified significant limitations. The chosen object detection model was found to be error-prone, and the overall system throughput is fundamentally limited by the 2.23-second autofocus operation. This work, therefore, provides both a complete, functional platform for autonomous screening and a clear, data-driven path for its future optimization. The integration of a hybrid brightfield imaging modality, as proposed in our discussion, will directly address these bottlenecks by enabling the use of both faster focus metrics and more accurate, pre-trained segmentation models, paving the way for this system's adoption in large-scale biological studies.

\bibliographystyle{IEEEtran}
\bibliography{references}

\end{document}